\documentclass[preprint]{revtex4}

\usepackage{graphicx}
\usepackage{dcolumn}
\usepackage{bm}

\usepackage{amsmath}
\usepackage{amssymb}
\usepackage{url}
\usepackage{subfigure}
\usepackage{algorithm}
\usepackage{algorithmic}
\usepackage{enumerate}
\usepackage{slashbox}
\usepackage{mathrsfs}

\begin{document}


\title{Effectiveness of Rapid Rail Transit System in Beijing}

\author{Hui-Min Cheng}
\affiliation{School of Statistics and Mathematics, Central University of Finance and Economics, P.R.China}
\author{Yi-Zi Ning}
\affiliation{School of Statistics and Mathematics, Central University of Finance and Economics, P.R.China}

\author{Xiaoke Ma}
\affiliation{School of Computer Science and Technology, Xidian University, P.R.China}

\author{Zhong-Yuan Zhang}
\email{zhyuanzh@gmail.com}
\affiliation{School of Statistics and Mathematics, Central University of Finance and Economics, P.R. China}

\date{\today}

\begin{abstract}
The effectiveness of rapid rail transit system is analyzed using tools of complex network for the first time. We  evaluated the effectiveness of the system in Beijing quantitatively from different perspectives, including descriptive statistics analysis, bridging property, centrality property, ability of connecting different part of the system  and ability
of disease spreading. The results showed that the public transport of Beijing does benefit from the rapid rail transit lines, but there is still room to improve. The paper concluded with some policy suggestions regarding how to promote the system. This study offered significant insight that can help understand the public transportation better. The methodology can be easily applied to analyze other urban public systems, such as electricity grid, water system, to develop more livable cities.
\end{abstract}

\pacs{Valid PACS appear here}
\maketitle

\section{\label{Introduction}Introduction}
Network science is deeply rooted in real applications, and there is strong emphasis on empirical data. Actually, The world is full of complex systems, such as organizations with cooperation among individuals, the central nervous system with interactions among neurons in our brain, the ecosystem, etc. It has been one of the major scientific challenges to describe, understand, predict, and eventually take good control of complex systems.
Indeed, the complex system can be naturally represented as a network that encodes the interactions between the system's components, and hence network science is at the heart of complex systems\cite{barabassi}. In general, there are three main aspects in the research of complex networks including (1) evolution of networks over time\cite{erdds1959random,gilbert1959random}, (2) topological structures of networks, such as scale-free property and community structures\cite{barabasi1999emergence,Girvan02}, (3) and role of the topological structures, such as how to influence spreading on networks\cite{newman2002spread,nematzadeh2014optimal}.
This paper focuses on the latter two ones.

During the last few years, more and more attentions have been paid to transportation network. Most of the works are about the topological characteristics of the transportation networks, including statistical analysis\cite{statistical,airportstatistical}, community structures\cite{de2013community}, effectiveness \cite{latora2001efficient}, centrality\cite{guimera2005worldwide,derrible2012network}, etc. Others are about the functions of the networks including robustness\cite{derrible2010complexity},  facilitating travel\cite{zhang2015evaluation}, epidemic spreading\cite{colizza2006role}, economy\cite{cho2001integrating}, etc.

Rapid rail transit system (RRTS for short) is an important part of public transportation. However, the cost of RRTS is usually very high, and few attempts have been made to evaluate its effectiveness in one city\cite{jarboui2012public}. To the best of our knowledge, this paper is the first time to quantitatively analyze the effectiveness using tools of complex network. We represent the Beijing transportation system as an unweighted directed network, and the main contributions are fourfold: (a) The transportation network has small world property. (b) Different from its counterparts in foreign cities, the transportation system has a high assortativity coefficient, which reduces the robustness of the entire system and may lead to traffic congestion. (c) The degree of the dependency on RRTS varies with different regions, and the benefit of different
regions from the system is gradually decreased from the
north to the south. (d) RRTS promotes the spread of communicable diseases.

The rest of the paper is organized as follows: Sect. \ref{cons} described in detail the spatial distribution pattern of the transport stations, and how the network was built. The  descriptive statistics of the data was also presented. From Sect. \ref{local} to Sect. \ref{SI}, we analyzed the effectiveness of Beijing RRTS from different perspectives, including bridging property, centrality property, ability of connecting different part of the system, ability of disease spreading. Finally, Sect. \ref{conclusion} concluded.

\section{\label{cons}Network Construction and description of Public Transport System}
\begin{figure*}
\hspace*{-10mm}
\includegraphics[height=70mm,width=140mm]{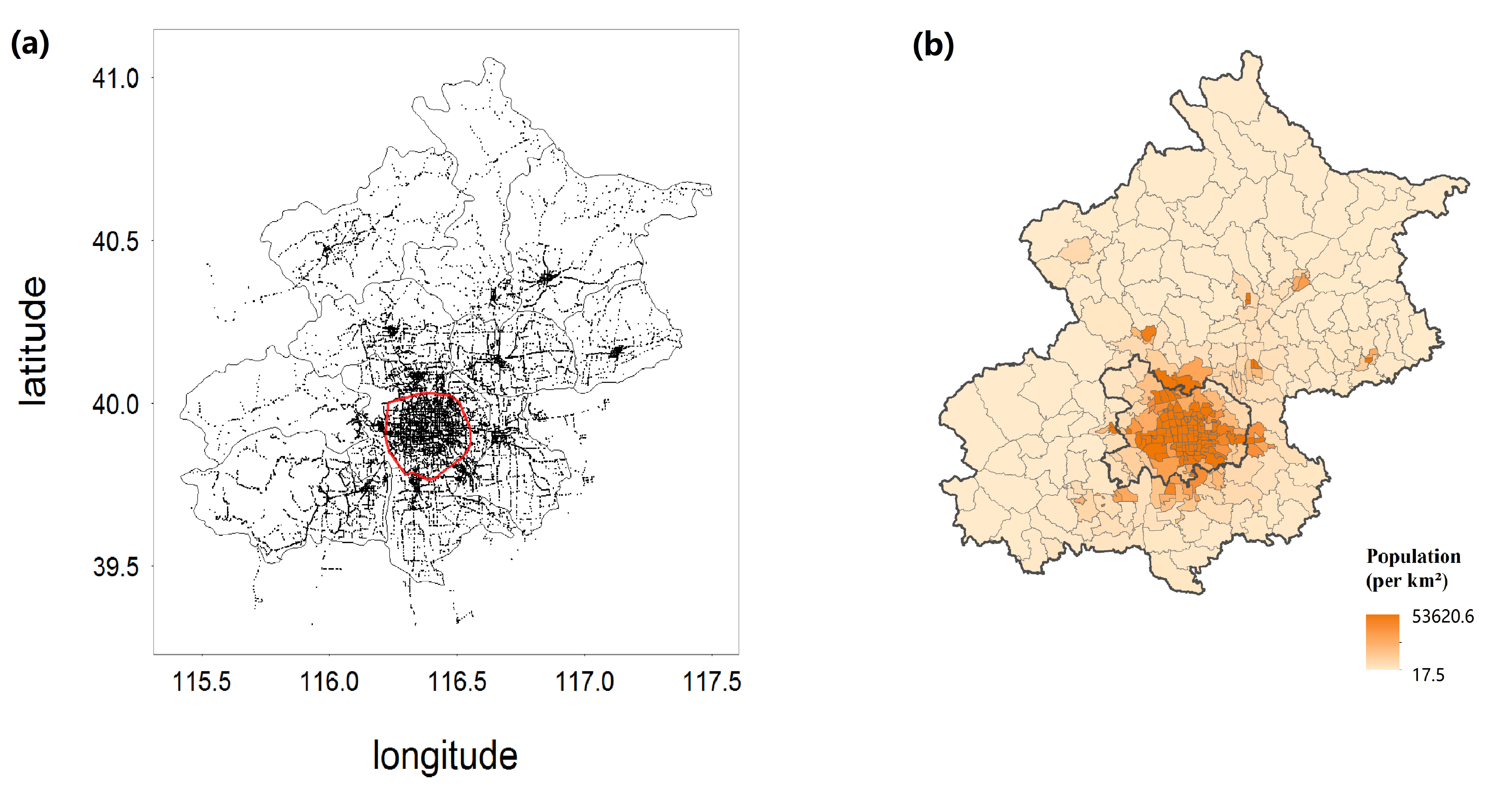}
\caption{(a) The spatial distribution of the public transport stations of Beijing, and (b) the spatial pattern of the population density. The red circle line in panel (a) shows the 5th ring road of Beijing.}\label{Fig:04}
\end{figure*}

Beijing is located in northern China, and its terrain is high in the northwest and is low in the southeast. It is the capital city of the People's Republic of China, and is governed as a direct-controlled municipality under the national government with 16 urban, suburban, and rural districts. It is the world's third most populous city proper, and is the political, economic and cultural center of China. The city spreads out in concentric ring roads, and the city's main urban area is within the 5th ring road.  We collected the public transport data of Beijing including city buses, trolley buses and rapid rail transit. Firstly, we graphically displayed the spatial distribution of the stations to find out the general pattern of Beijing public transportation system. As is shown is Fig. \ref{Fig:04}, The spatial pattern does match with the terrain of Beijing, and also with the spatial pattern of the population density of Beijing\cite{dong2014assessing}: (1) Overall, the coverage percentage of the public transport stations and the population density are all increased gradually from the northwest to the southeast, and are all relatively higher within the 5th ring road. (2) There are more stations in the areas, outside the 5th ring, with higher population densities.  The above analysis also reflects the accuracy of the data collected, making our further analysis and the conclusions more reliable.

The public transport system of Beijing can be naturally represented as a unweighted directed network $\mathscr{G}_1$, where the nodes are the stations, and the directed edge from node $i$ to node $j$ means that there is at least one route in which station $j$ is the successor of the station $i$ or the distance between them is less than $250$ meters (\texttt{m} for short).

Table \ref{Tab:05} gathers the fundamental descriptive statistics of the transportation networks with and without rapid rail transit stations, and of the rapid rail transit network, denoted by $\mathscr{G}_1$, $\mathscr{G}_2$ and $\mathscr{G}_3$ respectively, including number of nodes $N$ and edges $m$, median of in-degrees $I$ and out-degrees $O$, averaged shortest path distance $p$, clustering coefficient $c$\cite{wasserman1994social}
\footnote{The clustering coefficient of a network  is simply the ratio of the triangles and the connected triples in it. For directed network the direction of the edges is ignored.}, and assortativity coefficient $r$\cite{newman2002assortative}
\footnote{The assortativity coefficient of a directed and connected network is simply the Pearson correlation coefficient of degrees between pairs of linked nodes, and is defined as:
$$
r = \sum\limits_{jk}\frac{\left[jk(e_{jk}-q_{out,j}q_{in,k})\right]}{\sigma(q_{in})\sigma(q_{out})}
$$
where $q_{out,i}=\sum\limits_je_{ij}$, $q_{in,i}=\sum\limits_je_{ji}$, $\sigma(q_{out})$ and $\sigma(q_{in})$ are the standard deviations of $q_{out}$ and $q_{in}$, respectively.
$r$ is between $-1$ and $1$, and is used to measure whether nodes tend to be connected with other ones with similar degrees.}.
We also generated the randomized degree-preserving counterpart of the transportation network $\mathscr{G}_1$ for comparison, denoted by $\mathscr{G}_4$.

From the table, one can observe that: (1) The averaged shortest path distance $p$ of the transportation network $\mathscr{G}_1$ is comparable with $\log N \approx11$, and is also comparable with that of its randomized counterpart $\mathscr{G}_4$. The clustering coefficient of the network $\mathscr{G}_1$ is significantly larger than $\mathscr{G}_4$. The above two points show that the transportation network is not random, and is typically a small world network\cite{watts1998collective}. (2) The  clustering coefficient of the rapid rail transit network $\mathscr{G}_3$ is 0, meaning that there is no triangles in the network, as expected. (3) The averaged shortest path distance of network $\mathscr{G}_1$ is half station shorter than that of network $\mathscr{G}_2$, meaning that the public transport of Beijing does benefit from the rapid rail transit lines. (4) An interesting observation is that the assortativity coefficient of the public transportation network is high, which is very different from its counterpart \cite{statistical}. Recent studies showed that high assortativity within a single network decreases the robustness of the
entire system (network of networks) \cite{zhou2012assortativity}, which may lead to traffic congestion.

\begin{table}
\caption{Statistical description of the constructed networks. $\mathscr{G}_1$ means the network of Beijing public transport system; $\mathscr{G}_2$ means the network without rapid rail transit stations; $\mathscr{G}_3$ means the rapid rail transit network; $\mathscr{G}_4$ means the randomized degree-preserving network of $\mathscr{G}_1$. $N$ is the number of nodes; $m$ is the number of edges; $I$ is the median of in-degrees; $O$ is the median of out-degrees; $p$ is averaged shortest path distance; $c$ is clustering coefficient; $r$ is assortative coefficient. Note that the sum of $N$ of $\mathscr{G}_2$ and $\mathscr{G}_3$ is larger than the node number of $\mathscr{G}_1$ because there are two stations shared by $\mathscr{G}_2$ and $\mathscr{G}_3$.}
\centering\scriptsize
\begin{tabular}{c | c c c c c c c}\hline\hline
ID\hspace{2mm}  & \hspace{1mm} $N$ \hspace{1mm} & \hspace{1mm} $m$ \hspace{1mm} & \hspace{1mm} $I$ \hspace{1mm} & \hspace{1mm} $O$ \hspace{1mm} & \hspace{1mm} $p$ \hspace{1mm} & \hspace{1mm} $c$ \hspace{1mm} & \hspace{1mm} $r$ \hspace{1mm} \\\hline
 $\mathscr{G}_1$\hspace{2mm} & 72134 & 1486834 & 16 & 16  & 24.48 & 0.81 & 0.90 \\
 $\mathscr{G}_2$\hspace{2mm} & 71872 & 1475068 & 16 & 16 & 25.06 & \hspace{1mm}0.81+ \hspace{1mm}& 0.91 \\
 $\mathscr{G}_3$\hspace{2mm} & 264 & 588      & 2  & 2  & 14.22 & 0    & 0.07 \\
 $\mathscr{G}_4$\hspace{2mm} & -- & -- & -- & -- & 5.43 & 0.57 & --\\
 \hline\hline
\end{tabular}\label{Tab:05}
\end{table}
\section{\label{local}RRTS Lines have highER Local Bridge values}

To evaluate the local effectiveness of RRTS, we compared the local bridge values of the connected rapid rail transit stations, and those of the rest edges in the network $\mathscr{G}_1$. The local bridge value of a directed edge from $i$ to $j$ is the length of the shortest route from $i$ to $j$ after deleting the edge \cite{easley2010networks}.
For each connected nodes $i$ and $j$, the shortest route is obviously $i\rightarrow j$ (or $j\rightarrow i$), and the length is $1$. If we delete the edge, the length will be increased.
A longer route indicates that the edge is more powerful for connecting different parts of the network and is more important for the convenience of traffic.

The average of the local bridge values of the connected rapid rail transit stations is $4.81$, and that of the rest is $2.09$. We ran the independent samples T test, and the  $p-$value is less than $2.2\times10^{-16}$, meaning that the difference between the average values has statistical significance.

\begin{figure}
\centering
\includegraphics[height=60mm,width=80mm]{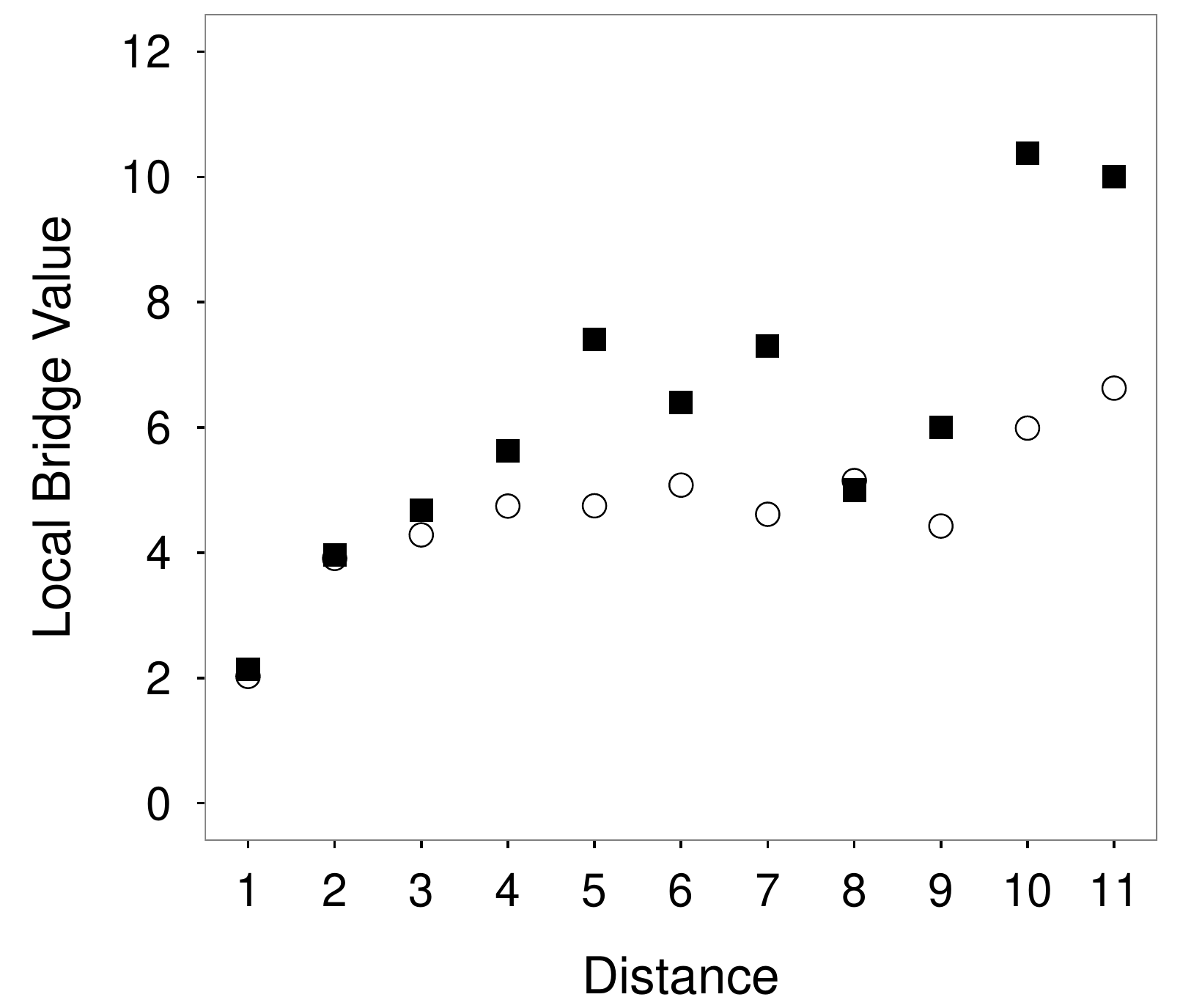}
\caption{The local bridge value versus the distance of connected nodes. The black squares are the averaged values of the connected rapid rail transit stations, and the white circles are the averaged values of the rest. The numbers on the x-axis mean intervals of distances: 1 means $0\texttt{m}-500\texttt{m},$ 2 means $500\texttt{m}-1000\texttt{m}$, $\cdots$, 11 means $5000\texttt{m}-5500\texttt{m}$.}\label{Fig:01}
\end{figure}

The local bridge value is obviously effected by the distance of the connected nodes. The longer the distance, the larger the bridge value. To keep things fair, we equally divided the range of distances of connected nodes into a series of intervals, i.e., $0\texttt{m}-500\texttt{m},$ $500\texttt{m}-1000\texttt{m}$, $\cdots$, $5000\texttt{m}-5500\texttt{m}$, and calculated the averaged local bridge values of the connected nodes fallen into each interval. From Fig.\ref{Fig:01}, one can observe that: (1) The local bridge values are increased when the distances between the stations are increasing, as expected; (2) The values of the edges connecting the rapid rail transit stations are consistently higher than those of the rest, especially when the distance is large, indicating that the stations connected by the rapid rail lines have fewer common neighbors, and  are more important for connecting different parts of the network \cite{granovetter1973strength}.
\section{\label{between}RRTS Lines have higher Centrality}
To evaluate the centrality of the rapid rail transit stations, we calculated the betweenness values \cite{freeman1977set} and the closeness values \cite{bavelas1950communication,sabidussi1966centrality} of stations in the network $\mathscr{G}_1$.

In a connected network, the betweenness centrality of node $v$ is defined by
$$
\sum\limits_{i,\,j}\frac{g_{ivj}}{g_{ij}},\hspace{5mm} i\neq j,i\neq v,j\neq v,
$$
where $g_{ij}$ is the total number of shortest paths from node $i$ to $j$, and $g_{ivj}$  is the number of those paths that pass through $v$.

Betweenness centrality is introduced as a measure for quantifying the control of a node on the communication between other nodes in a complex network.

In a connected and directed network, the closeness\_in centrality of node $v$
 is defined as the inverse of the average length of the shortest paths from all the other vertices in the graph:
$$
\frac{1}{\sum\limits_{i} d(i,v)},\hspace{5mm} i \neq v,
$$
and the closeness\_out centrality is defined as that to all the other ones:
$$
\frac{1}{\sum\limits_{i} d(v,i)},\hspace{5mm} i \neq v,
$$
where $d(i,v)$ is the shortest path length from node $i$ to $v$ in the directed network.
If there is no (directed) path between node $v$ and $i$, then the total number of nodes is used in the formula instead of the path length.

A larger closeness value of node $v$ means that the total distance to/from all other nodes from/to $v$ is lower, and the node $v$ is in the middle of the network.

We listed the top ranked $20$ stations based on different centrality measures in Table \ref{Tab:02}, and the stations that are appeared in all of the three lists are underlined. One can observe that: (1) Generally, the two centrality metrics are positively correlated. (2) Some stations are the exceptions. They have higher betweenness but lower closeness, or conversely, have lower betweenness but higher closeness.

\begin{figure}
\hspace*{-10mm}
\includegraphics[height=50mm,width=160mm]{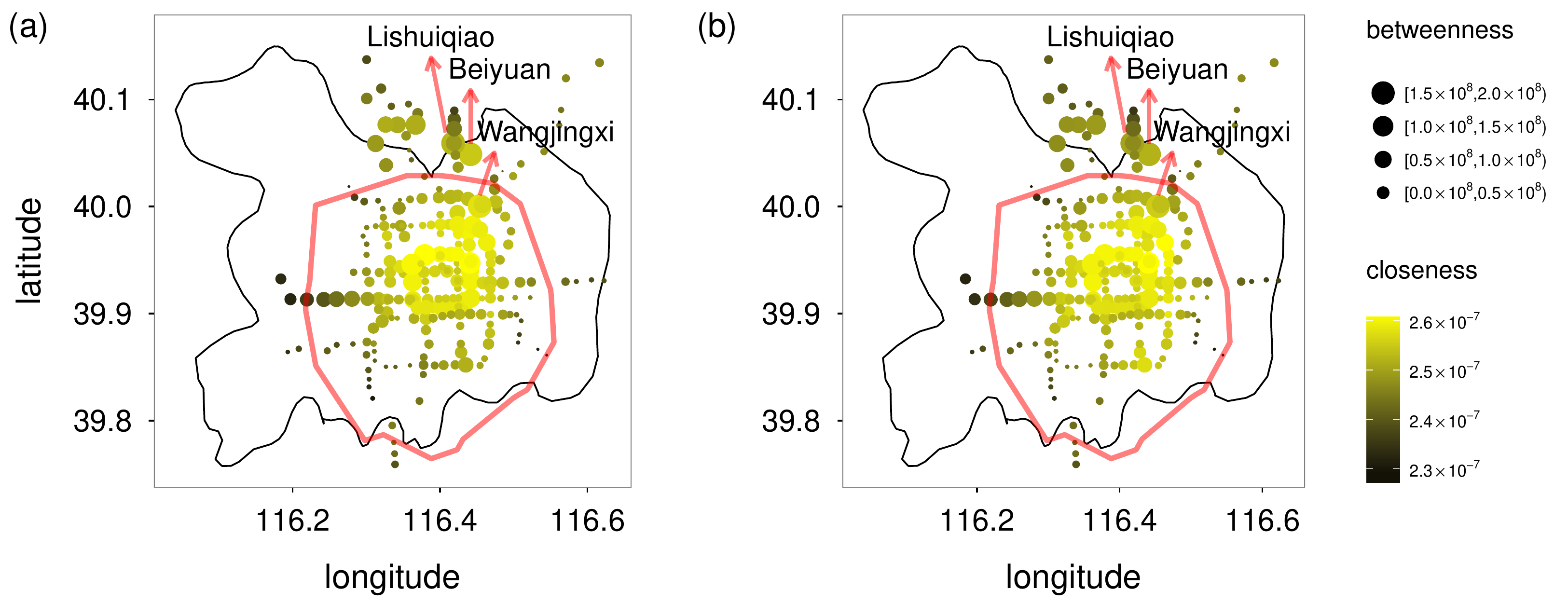}
\caption{Betweenness versus (a) closeness\_in and (b) closeness\_out of the rapid rail transit stations. The bigger the circles, the larger the betweenness values. The more bright the color, the larger the closeness values. The stations in the central region have higher closeness, as expected. The stations in northern Beijing have higher betweenness but lower closeness, such as Lishuiqiao, Wangjingxi and Beiyuan, indicating that they monopolize
the connections from a small number of stations to many others.}\label{Fig:06}
\end{figure}

In order to further study the positions of the stations in the network, we drew RRTS on the Beijing map, as is shown in Fig. \ref{Fig:06},
from which, one can observe that: (1) The stations in the central region of Beijing have higher closeness values, as expected. (2) The stations in the northern Beijing have higher betweenness values, and several typical stations are marked on the map. (3) The betweenness values decrease gradually from the north to the south, indicating that northern Beijing is more dependent on RRTS.

\begin{figure}
\hspace*{-15mm}
\includegraphics[height=50mm,width=170mm]{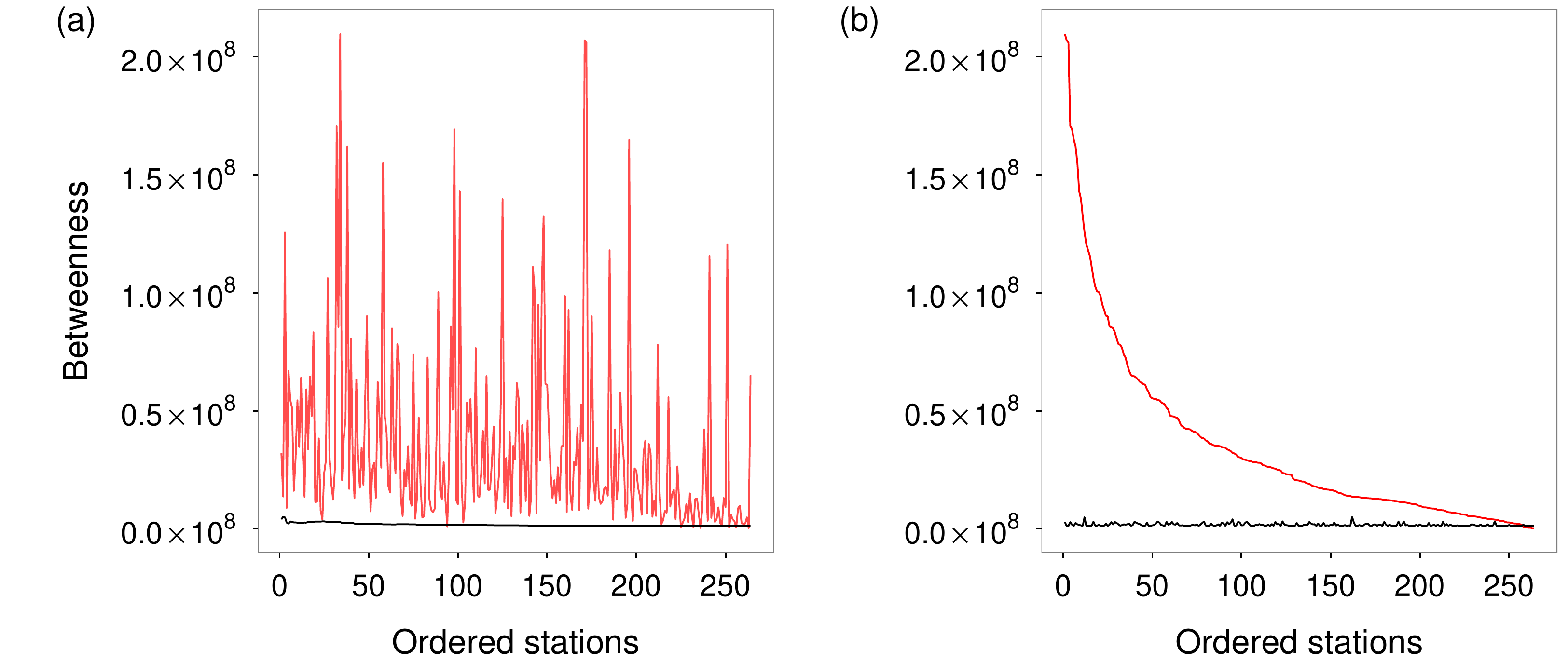}
\caption{Red line: betweenness values of the rapid rail transit stations ordered by (a) the degrees, and (b) the betweenness values; Black line: averaged betweenness values of the other stations with comparable degrees.}\label{Fig:05}
\end{figure}

We also compared the betweenness of the rapid rail transit stations with the other ones with comparable degrees, which is summarized in Fig. \ref{Fig:05}. One can see that: (1) The betweenness values of rapid rail transit stations are generally larger than their counterparts, indicating that RRTS is really important for bringing convenience to the transportation of citizens. (2) The betweenness values of rapid rail transit stations are gradually decreased with the decreasing of degree.

\begin{table}
\caption{The top 20 rapid rail transit stations with the highest betweenness values and the highest closeness values. The underlined stations are those appeared in all of the lists.}
\centering\scriptsize
\begin{tabular}{c || l}\hline\hline
  Centrality & Top ranked stations  \\\hline
 Betweenness & Lishuiqiao, Wangjingxi (line 13), Beiyuan, \underline{Jishuitan}, \underline{Dongzhimen (line 2)}, \\
 & \underline{Xizhimen}, Congwenmen, Qianmen, Huoying, Beijingzhan,\\
 &  \underline{Chaoyangmen}, \underline{Fuchengmen}, \underline{Jianguomen}, \underline{Shaoyaoju},\\
 & \underline{Dosishitiao},  \underline{Chegongzhuang}, \underline{Sanyuanqiao}, Wangjingxi (line 15), \\
 & \underline{Huixinxijienankou}, Xierqi \\\hline
 Closeness\_in & \underline{Jishuitan}, \underline{Dozhimen (line 2)}, Yonghegong, \underline{Xizhimen}, Guloudajie, \\
 & Andingmen, Dongzhimen (line 13), \underline{Chegongzhuang}, Guangximen, Shaoyaoju, \\
 & \underline{Chaoyangmen}, \underline{Dongsishitiao}, \underline{Fuchengmen}, \underline{Sanyuanqiao}, \\
 & \underline{Huixinxijienankou}, Hepingxiqiao, \underline{Jianguomen}, Beijingzhan, \\
 & \underline{Shaoyaoju}, Taiyanggong \\\hline
 Closeness\_out & \underline{Dongzhimen (line 2)}, \underline{Sanyuanqiao}, Yonghegong, Nongyezhanlanguan,\\
 & \underline{Jishuitan}, Guloudajie, Dongzhimen (line 13), \underline{Fuchengmen}, Andingmen,\\
 & Liangmaqiao, \underline{Xizhimen}, \underline{Chegongzhuang}, Hepingxiqiao, \underline{Shaoyaoju}, \\
 & \underline{Dongsishitiao}, \underline{Chaoyangmen}, \underline{Huixinxijienankou}, Fuxingmen,\\
 & Taiyanggong, \underline{Jianguomen} \\
\hline\hline
\end{tabular}\label{Tab:02}
\end{table}
\section{\label{community}rrts Lines Make Travel More Convenient}
\begin{figure*}
\includegraphics[height=60mm,width=120mm]{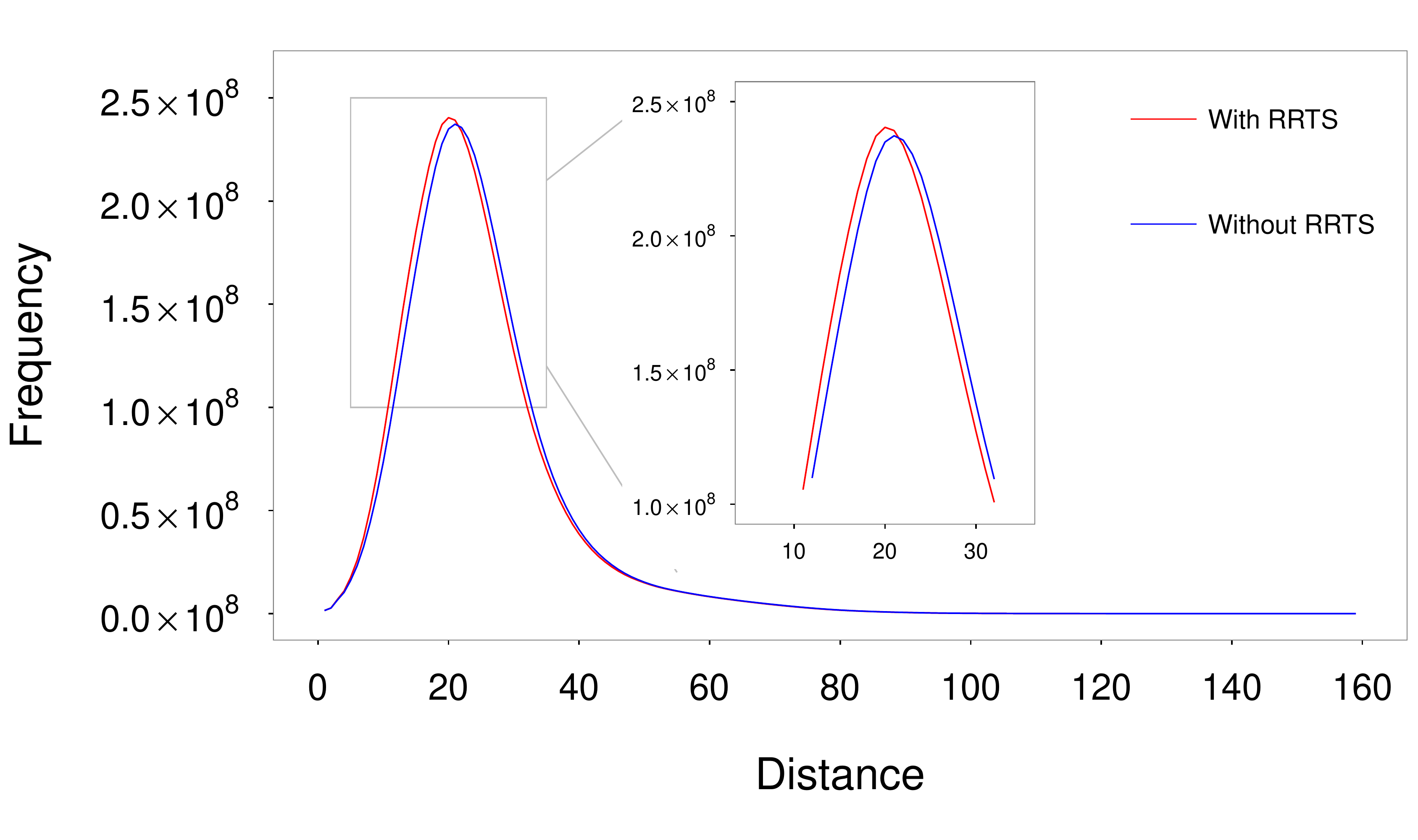}
\caption{Frequency distribution of shortest path distances in the network with rapid rail transit stations (red line) and that without them (blue lines). Inset is the partial enlarged plot.}\label{Fig:08}
\end{figure*}

Results of Sect. \ref{cons} indicate that, on average, one's traveling distance is just half station longer if he/she does not use RRTS. Fig. \ref{Fig:08} is the frequency distribution of shortest path distances in Beijing's transportation network, which shows that the shift to left is very small. Both results suggest that the benefit of RRTS is limited. But this is not the full story. The following analysis shows that RRTS makes the connections of different regions in Beijing more efficient.

\begin{figure*}
\hspace{-10mm}
\includegraphics[height=90mm,width=140mm]{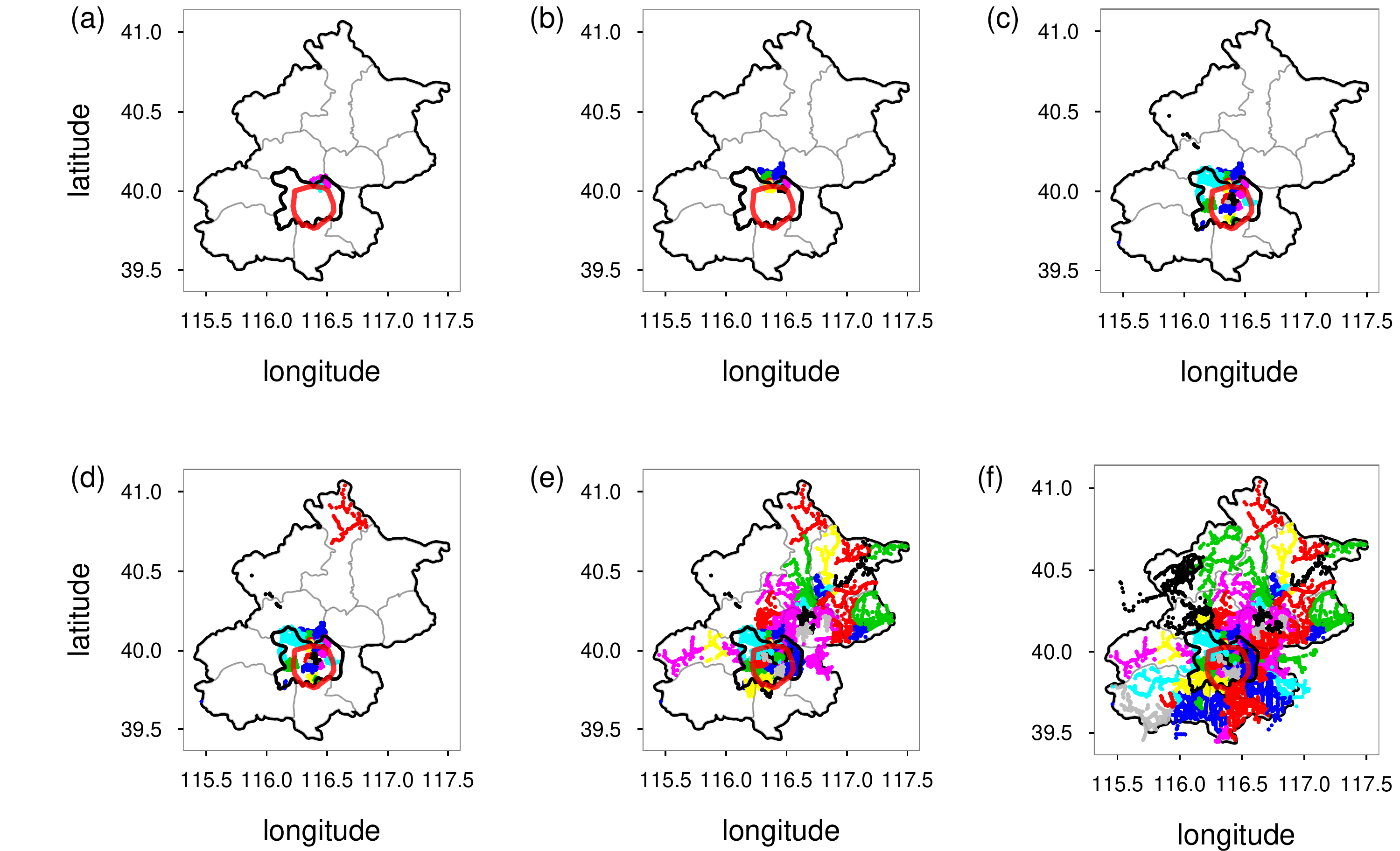}
\caption{Community structures of the public transportation network without rapid rail transit stations. Communities are displayed on the map in descending order as to their reception of benefits from RRTS from the most benefitted to least benefitted.}\label{Fig:09}
\end{figure*}

Firstly, we detected the community structures in the transportation network without rapid rail transit stations using the fast greedy modularity optimization algorithm\cite{clauset2004finding}, and the network was partitioned into $94$ non-overlapping communities. A community in the network is a set of nodes that are densely interconnected but loosely connected with the rest of the network\cite{Girvan02}. Secondly, we calculated the distances among the communities with and without rapid rail transit stations, i.e., the distance from community $i$ to $j$ is the average of the pairwise distances from the nodes in community $i$ to those in community $j$, and we compared the difference.
The overall results are shown in Fig. \ref{Fig:09}, from which one can see that: (a) The benefit of different regions from RRTS is different, and is gradually decreased from the north to the south,
which is in accordance with the results of Sect. \ref{between}.
(b) The region of Lishuiqiao benefits the most from RRTS, and one's averaged distance of traveling to the rest of Beijing is $2.93$ stations shorter. Take for example the route from the Xichengjiayuan station to the Shahegaojiaoyuan station, which is actually the last author's commute to work. Without RRTS, the distance is 24 stations, and is 6 stations shorter with the system. (c) The region of Yanqing, which is in northwest Beijing, benefits the least, and one's averaged travelling distance is only $0.09$ stations shorter.

\section{\label{SI}RRTS lines promote spread of disease}
Finally, we evaluated the efficiency of RRTS for disease spreading. We adopted the susceptible-infected (SI) disease model, which is suitable for simulating the beginning stage of the diffusion and can be formulated as follows: at time $0$, some nodes are randomly selected and are set to be infected. During the disease diffusion,  each node has two possible states: S (susceptible) and I (infected). At time $t+1$, susceptible node can become infected with possibility $\lambda$ if it has an infected neighbor at time $t$, and with possibility $1-(1-\lambda)^k$ if it has $k$ infected neighbors at time $t$. We set $\lambda$ to be $0.4$.

\begin{figure*}
\hspace*{-10mm}
\includegraphics[height=40mm,width=170mm]{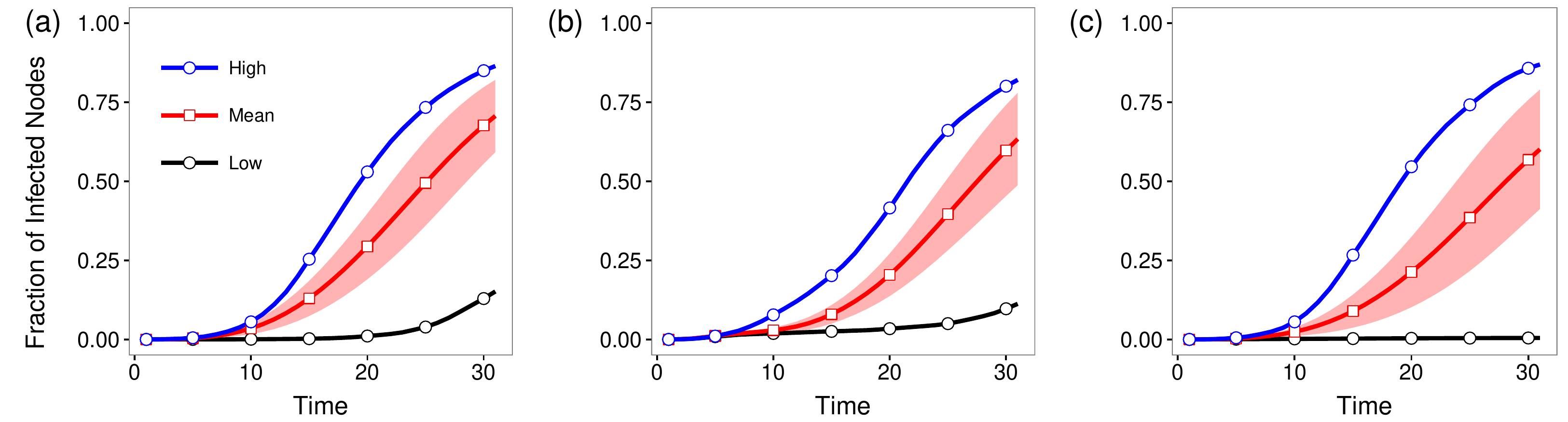}
\caption{Time series of the averaged fraction of infected nodes with standard deviation in the public transport network of Beijing. ``High'' means the diffusion curve of the station with the highest spreading capacity at the time step $30$, ``Mean'' means the averaged fraction, and ``Low'' means the curve of that with the lowest spreading capacity at the time step $30$. The results are averages of five trials. The initial infected nodes are (a) the rapid rail transit stations; (b) the nodes with the highest degrees in the network; (c) the nodes with the degrees comparable with that of the rapid rail transit stations.}\label{Fig:03}
\end{figure*}

There are $264$ rapid rail transit stations. We also selected $264$ stations with the highest degrees and randomly selected $264$ stations with the degrees comparable with those of the rapid rail transit stations as the candidates for comparison. At each time, for the initial condition of diffusion, we selected one station as the infected spreader. The evolution of the infected nodes over time is shown in Fig. \ref{Fig:03}. The results are averages of five trials.

From Fig. \ref{Fig:03}, one can observe that: (1) The rapid rail transit stations are in general more efficient with smaller deviation for disease diffusion. (2) The spreading capability of stations with the highest degrees is comparable with that of the rapid rail transit stations. (3) For the nodes whose degrees are comparable with those of the rapid rail transit stations, the spreading capacity is lower. (4) Rapid screening is reasonable during the outbreak of infectious diseases for detecting people with elevated body temperatures.
\section{Conclusions and Future work}\label{conclusion}
In this paper, the effectiveness of RRTS was evaluated using complex network analysis theory.  We represented Beijing public transportation system as an unweighted directed network, and evaluated the properties of RRTS from different perspectives, including descriptive statistics analysis, bridging property, centrality property, ability of connecting communities in the system, and ability of disease spreading.
In summary: (1) The public transportation system has small world property. (2) The rapid rail transit lines are weak ties, have higher centrality, and are important for connecting different communities in the transportation system, making travelling more convenient. (3) As a byproduct, the rail lines promote the spread of disease.
Based on the findings that the transportation system has high assortativity, reducing the robustness of the entire system, and people in northern Beijing is more dependent on RRTS than that in the rest parts of Beijing, our policy suggestions include more consideration to the rapid rail transit construction in the south and to the ground public transportation construction in the north, especially in the area of Lishuiqiao, construction of more lines connecting the nodes with different degrees to reduce assortativity, and body temperature rapid screening during the epidemics.

Based on our works, there are several interesting problems for future work, including  traffic early warning by combining data from other resources, such as congestion prediction, helping plan the rapid rail route and station locations to improve the efficiency of the transportation system, and comprehensive comparison of the topological structure and statistical properties of the transportation systems in different scale cities.
\begin{acknowledgments}
\end{acknowledgments}
\bibliography{reference}
\end{document}